\documentstyle[12pt]{article}
\begin{document}
\newcommand{\be}{\begin{equation}}
\newcommand{\ee}{\end{equation}}          
\begin{center}
{\Large\bf Glueballs in the String Quark Model} \\
\vspace{0.3cm}
{\large\bf L. D. Soloviev }\\
\vspace{0.3cm}
{\it Institute for High Energy Physics, 142284, Protvino, Moscow
region}
\end{center}
\vspace{0.5cm}
{\small It is shown that the eigenstates of the quantized simplest
closed
(elliptic) Nambu -- Goto string, called glueballs, have quantum
numbers $I^Gj^{PC}=0^+j^{++}$. Lightest glueballs have spins $j=0,1$
and $2$ and the same mass $1500\pm20 ~MeV$. They correspond to
$f_0(1500)$, $f_1(1510)$ and $f_2(1565)$- mesons. Next glueballs have
$j=0,1,2,3,4$ and the same mass $2610\pm20 ~MeV$. The slope of the
glueball Regge trajectories is twice as small as for $q\bar
q$-mesons.
The intersept of the leading glueball trajectory --- the pomeron
Regge
trajectory ---is $1.07\pm0.03$.}

\vspace{1cm}
\section {Introduction}

Overwhelming majority of known mesons can be described as eigenstates
of the quantized straight-line Nambu -- Goto string with the Dirac
quarks at the string ends [1]. The string approximately accounts for
the nonperturbative contribution of the gluon field at large
distances and is characterized by a string tension parameter. The
quarks are characterized by their current masses. All theese
parameters, together with phenomenological parameters describing
nonstring short-range gluon contribution, are determined from
comparison of the theoretical and experimenral meson mass spectra
for mesons lying on the leading Regge trajectories.

Since this approach has appeared to be successful for the
quark-antiquark mesons, it seems natural to generalized it to the
simplest {\it closed} strings with the aim to describe the gluon
meson
states called glueballs.

The classical and quantum description of the simplest closed
(elliptic) Nambu -- Goto string was done in Ref. [2]. In the present
paper a nonstring short-range gluon contribution is introduced and
possible iterpretation of the results of Ref. [2]is considerd
together with the $q\bar q$-meson analysis of Ref. [1].

In Sec. 2 the results of Ref. [2] are obtained by the 1-form method
of Ref. [3] which is simpler than the Dirac brackets method of Ref.
[2]. Properties of the classical solution, its stability in
particular, are also considered. The solution is not stable when the
string mass is zero. This value of mass must be excluded when
quantizing the string (otherwise the quantization would lead to a
tachion). In Sec. 3 the canonical quantization, glueball wave
functions and glueball Regge trajectories are considered. Comparison
with experiment allows one to identify the lowest glueball states,
to fix the only free parameter of the model and to calculate the
model predictions: masses and quantum numbers of higher glueballs and
the pomeron Regge trajectory. In conclusion, the obtained results are
briefly discussed.

\section {Classical glueball model }

A closed string is described by a 4-vector $x^{\mu}(\sigma, \tau)$,
which depends on relativistic scalars characterizing position of
points on the string ($\sigma$) and evolution of the string ($\tau$); 
$0 \le \sigma \le 2\pi$, $x^{\mu}(0, \tau)=x^{\mu}(2\pi, \tau)$.
Prime and dot denote partial derivatives with respect to $\sigma$ and
$\tau$,
respectively. The Nambu -- Goto Lagrangian
\be
{\cal L}=-a\int_{0}^{2\pi}((x'\dot x)^2-x'^2\dot x^2)^{1/2}\,d\sigma,
\ee
where the string tension parameter $a$ is known from the $q\bar q$-
meson analysis [1], can be represented in the Hamiltonian form [4]
\be
{\cal L}=-\int_{0}^{2\pi}p\dot x\,d\sigma - \int_{0}^{2\pi}
(h_1 (a^2x'^2 + p^2) + h_2 px')\,d\sigma.
\ee
Here the first term determines the Poisson brackets of the components
of the string coordinate $x$ and the conjugate momentum density $p$
[3] and the second term is the string Hamiltonian which, due to the
Lagrangian symmetry, is a linear combination of the constraint
functions  ($h_i$ are Lagrange multipliers).

Let us consider the simplest (elliptic) cofiguration of the closed
string
\be
x(\sigma, \tau) = r(\tau) + q_1(\tau)\cos \sigma + q_2(\tau)\sin
\sigma
\ee
and use parametrization in which the momentum density has the same
configuration
\be
p(\sigma,\tau)=\pi^{-1}(2^{-1}P(\tau)+\pi_1(\tau)\cos \sigma
+ \pi_2(\tau)\sin \sigma).
\ee
The orthonormal parametrization, for instance, in which $p=a\dot x$,
or parametrization in which $x'\dot x=0$ and $\dot x^2/x'^2$ does not
depend on $\sigma$, possess this property.

Putting (3) and (4) into (2), we get constraints which correspond
both to the Lagrangian symmetry and to the choice of parametrization
\be 
{\cal L}=-P\dot r-\pi_1\dot q_1-\pi_2\dot
q_2-\sum_{i=1}^{10}c_i\phi_i,
\ee
where $c_i$ are proportional to integrals over $d\sigma$ of
$h_{1,2}$, multiplied by $1, \cos k\sigma$ or $\sin k\sigma$, $k=1$
or $2$.
The constraints $\phi_i$ are
\be
\phi_1=a^2\pi^2(q_1^2+q_2^2)+2^{-1}P^2+\pi_1^2+\pi_2^2,
\ee
\be
\phi_2=\pi_1 q_2-\pi_2 q_1
\ee
\be
\phi_3=P\pi_1, ~~\phi_4=P\pi_2,~~\phi_5=Pq_2,~~\phi_6=Pq_1,
\ee
\be
\phi_7=a^2\pi^2(q_2^2-q_1^2)+\pi_1^2-\pi_2^2,~~\phi_8=-a^2\pi^2
q_1q_2+\pi_1\pi_2,
\ee
\be
\phi_9=\pi_1 q_2+\pi_2 q_1,~~\phi_{10}=\pi_2 q_2-\pi_1 q_1.
\ee
The 1-form in (5) determines the Poisson brackets of the introduced
variables. The nonzero brackets are
\be
\{P^{\mu},r^{\nu}\}=\{\pi_i^{\mu},q_i^{\nu}\}=g^{\mu\nu},~~i=1,2.
\ee
The constraints (8), (9), (10) are second kind constraints with
respect to theese brackets (due to the choice of parametrization)and
must be solved explicitly. To this end we first consider conserved
and parametrization-invariant string variables which have zero
brackets with the Hamiltonian in (5). Such variables are the total
string momentum $P$ and its nass $m=\sqrt{P^2}$ and the string spin
\be
J_{\mu}=\sum_{i=1,2}\epsilon_{\mu\nu\rho\sigma}v^{\nu}q_i^{\rho}
\pi_i^{\sigma},~~v^{\nu}=P^{\nu}/m.
\ee
$J_{\mu}$ has zero brackets with $P^{\nu}$ and with string Lorentz
scalars. The Poisson brackets of the spin with string (pseudo)vectors 
$Y$ are
\be
\{J_{\mu},Y_{\nu}\}=\epsilon_{\mu\nu\alpha\beta}v^{\alpha}Y^{\beta}.
\ee
It is remarkable that the elliptic string has one more conserved
parametrization-invariant pseudovector [2]
\be
W_{\mu}=\epsilon_{\mu\nu\rho\sigma}v^{\nu}((a\pi)^{-1}\pi_1^{\rho}
\pi_2^{\sigma}+a\pi q_1^{\rho} q_2^{\sigma}).
\ee
Its brackets with all constraints (6 -- 10) vanish and
\be
\{W_{\mu},W_{\nu}\}=\epsilon_{\mu\nu\alpha\beta}v^{\alpha}J^{\beta}.
\ee
Let us introduce so-called string {\it pseudospins}
\be
L_{1,2}=2^{-1}(J \pm W).
\ee
Their brackets are
\be
\{L_{i\mu},L_{j\nu}\}=\delta_{ij}\epsilon_{\mu\nu\alpha\beta}
v^{\alpha}L_i^{\beta},
\ee
\be
\{L_{i\mu},P_{\nu}\}=0.
\ee
Thus, the elliptic string has two independent conserved pseudospins
the sum of which is the string spin
\be
J=L_1 + L_2.
\ee
Now we can write down the solution of the constraints (8), (9) and
(10). Let us introduce the tetrad of vectors $e_{\alpha}^{\mu}(P)$,
$\alpha=0,a;~~a=1,2,3$
\be
e_{\alpha}e_{\beta}=g_{\alpha\beta},~~e_0=v,~~
\epsilon_{\mu\nu\rho\sigma}e_0^{\mu}e_a^{\nu}e_b^{\rho}
e_c^{\sigma}=\epsilon_{abc}
\ee
and expand the vectors $Y=q_i, \pi_i, L_i, J$, orthogonal to $P$,
with respect to the tetrad
\be
Y^{\mu}=e_a^{\mu}Y^a,~~~a=1,2,3.
\ee
We shall use the notations of 3-vectors for  a set $\{Y^a\}$ 
\be
\{Y^a\}=\vec Y,~~~\{\epsilon_{abc}n^b m^c\}=[\vec n,\vec m]
\ee
and so on. Then the solution of the constraints is
\be
\vec q_1=-\frac{m}{4\pi a}([\hat L_1,\vec n_1]
+[\hat L_2,\vec n_2]),~~~\vec \pi_1=\frac{m}{4}(\vec n_1+\vec n_2),
\ee
\be
\vec q_2=\frac{m}{4\pi a}(\vec n_1-\vec n_2),~~~\vec \pi_2=
\frac{m}{4}([\hat L_1,\vec n_1]-[\hat L_2,\vec n_2]),
\ee
where
\be
\hat L_i=\vec L_i/\mid\vec L_i\mid,
\ee
\be
\vec n_i^2=1,~~~\vec n_i\vec L_i=0.
\ee
Instead of 16 variables $q_i^{\mu}, \pi_i^{\mu}$ with 8 constraints,
we have introduced 12 variables $L_i^a, n_i^a$ with 4 constraints
(26), which can be easily solved, for instance, using spherical
coordinate systems.

Putting the solution (23 -- 26) into the Lagrangian (5), we get
\be
{\cal L}=-P\dot Z+\sum_{i=1,2}{[\vec L_i,\vec n_i]\dot{\vec n_i}}
-c_1(\sqrt{\vec L_1^2}+ \sqrt{\vec L_2^2}-\frac{P^2}{4\pi a})
-c_2(\sqrt{\vec L_1^2}-\sqrt{\vec L_2^2}),
\ee
where the coordinate conjugate to the total momentum is equal to
\be
Z_{\mu}=r_{\mu}+\frac{1}{2}\epsilon_{abc}e_a^{\nu}
\frac{\partial e_{b\nu}}{\partial P^{\mu}}J_c.
\ee
Nonzero brackets following from the 1-form in (27) are
\be
\{P^{\mu},Z^{\nu}\}=g^{\mu\nu},
\ee
\be
\{L_{ia},L_{ib}\}=\epsilon_{abc}L_{ic},~~~\{L_{ia},n_{ib}\}=
\epsilon_{abc}n_{ic}.
\ee
The constraints in (27) and the brackets (29), (30) were first
obtained in Ref. [2] by a different method.

Using stationary condition for (27), it is not difficult to obtain
the classical solution for the elliptic string. In the orthonormal
parametrization ($c_1=
1, c_2=0$ after variations)
\be
x(\sigma,\tau)=x_0+\frac{1}{2\pi a}P\tau + e_aq^a(\sigma, \tau),
\ee
\be
\vec q(\sigma,\tau)=\frac{m}{2\pi a}(-\sin \alpha 
\sin(\sigma+\sigma_0)\cos(\tau+\tau_0)\vec k 
+\cos(\sigma+\sigma_0)N\vec f(\tau)),
\ee
where $\vec k$ and $\vec f$  are unit orthogonal vectors:
$\vec k$  is a constant vector along the string spin and the vector
$\vec f(\tau)$ rotates around it. Introducing constant unit vectors
$\vec i$ and $\vec j$, orthogonal to $\vec k$ and to each other we
have
\be
\vec k=\vec J/|\vec J|=(\vec L_1+\vec L_2)/|\vec L_1+\vec L_2|,
\ee
\be
\vec i=(\vec L_2-\vec L_1)/|\vec L_2-\vec L_1|,~~~
\vec j=[\vec k,\vec i]
\ee
\be
\vec f(\tau)=N^{-1}(-\cos \alpha \sin(\tau+\tau_0)\vec i
+\cos(\tau+\tau_0)\vec j),
\ee
\be
N=(1-\sin^2\alpha \sin^2(\tau+\tau_0))^{1/2}.
\ee
The parameters $\alpha, \sigma_0$ and $\tau_0$ characterize the
initial conditions for the string: $\alpha$ is half of the angle
between constant vectors of pseudospins $\vec L_1$ and $\vec L_2$
having the same length and $\sigma_0$ and $\tau_0$ determine the
initial conditions of the unit vectors $\vec n_i$,
\be
\vec n_i=\vec n_{i1}\sin \tau- \vec n_{i2}\cos \tau,
\ee
in the coordinates $\vec i,\vec j,\vec k$
\be
\vec n_{i1}=(\sin\beta_i \cos\alpha,~-\cos\beta_i,~\sin\beta_i
\sin\alpha),~~\vec n_{i2}=(\cos\beta_i \cos\alpha,~\sin\beta_i,~
\cos\beta_i \sin\alpha),
\ee
\be
\tau_0=(\beta_1+\beta_2)/2,~~~\sigma_0=(\beta_1-\beta_2)/2.
\ee
Let us consider the motion of the string in the frame where it is at
rest as a whole,
$\vec P=0$. The laboratory time is
\be
t\equiv x^0-x_0^0=d\tau,~~~d=\frac{m}{2\pi a}.
\ee
The string behavior essentially depends on the pseudospins
$\vec L_1$ and $\vec L_2$, $|\vec L_1|=|\vec L_2|=L$. We have
\be
m^2=8\pi aL,~~~\vec J=\vec L_1+\vec L_2.
\ee
In case $\vec L_1=-\vec L_2$, the string spin is zero, the string
lies in the plane orthogonal to $\vec L_i$ and represents a
circumference with an oscillating radius, from maximal value $d$ to
zero and back.

In the general case, when the angle between $\vec L_1$ and $\vec L_2$ 
is in the limits $0<2\alpha<\pi$, the value of the string spin is
\be
J=\frac{m^2}{4\pi a}\cos\alpha.
\ee
The string represents an ellips with half-axises
\be
A=dN,~~~B=d\sin\alpha\cos(\tau+\tau_0)
\ee
(the large half-axis $A$ is orthogonal to the spin and the small one
is parallel to the spin) and rotates around the spin with the angular
velocity
\be
|d\vec f/dt|=d^{-1}\cos\alpha N^{-2}.
\ee
The half-axises are maximal: $A=d$ ¨ $B=d\sin\alpha$, and the instant
angular velocity is minimal: $d^{-1}\cos\alpha$, when the string is
in the plane orthogonal to the plane of $\vec L_i$. Rotating, the
ellips shrinks and accelerates. When it reaches the plane of $\vec
L_i$, it shrinks into a straight-line with half-length
$d\cos\alpha$. Its angular velocity at this moment is maximal and is
equal to the inverse of its half-length. Then the ellips expands and 
slows down and so on.

In the other extreme case, $\vec L_1=\vec L_2$, the string spin is
maximal
\be
J=\frac{m^2}{4\pi a}
\ee
(and twice as small as for an open straight-line string of the same
mass), the string is compressed into a straight-line with half-length
$d$ and rotates with a constant angular velocity $d^{-1}$.

To quantize the classical solution, one has to know the domain of its
stability. It is not difficult to show that the above solution is
stable for all values of the initial conditions except for $L=0$.
For $L=0$ the string reduces to a pointlike system with zero mass as
it is seen from (27), however for arbitrary small $L\ne0$ we have
system with small mass but with all degrees of freedom of the
elliptic string. Therefore quantization of our system has no meaning
for values of pseudospins close to zero (it would lead to a tachion).

\section {Quantization and comparison with experiment}

Canonical quantization of the model means replacement of the
variables $P,Z,\vec L_i$ and $\vec n_i$ by operators, their Poisson
brackets (29) and (30) by commutators  ($\{,\}\to -i[,]$) and the
constraints following from (27) --- by wave equations
\be
(\sum_{i=1,2}{\sqrt{\vec L_i^2}}-\frac{1}{4\pi a}P^2-a_0)\psi=0,
\ee
\be
(\sqrt{\vec L_1^2}-\sqrt{\vec L_2^2})\psi=0.
\ee
This quantum system is Poincar\'e-invariant because the operators of
linear and angular momenta, obtained from the corresponding classical
expressions, satisfy the Poincar\'e algebra [2].  $a_0$ has been
introduced into Eq. (46) to account for nonstring interaction at
small distances just as it was done for quark-antiquark mesons in
Ref. [1]. In general, $a_0$ could depend on  $\vec L_i$, but it can
not increase with $\vec L_i^2$, because the increasing contribution
comes from the string only (the main assumption of the model). Vacuum
fluctuations of the higher string modes could also contribute to
$a_0$. To a first approximation, let us take into account only a part
of $a_0$ which does not depend on  $\vec L_i$, that is we shall
consider $a_0$ as a constant to be determined from experiment. This
is the only unknown parameter of the model since the second
parameter, the string tension $a$, is known from the analysis of
quark-antiquark mesons [1]
\be
a=0,176 \pm 0,002~Ĵ'^2. 
\ee
Digressing from a factor in the wave function which describes the
motion of the string as a whole, we can write down the solution of
Eqs. (46), (47) in the representation where $P$ and $\vec n_i$ are
diagonal [2]
\be
\psi_{jMl}(\vec n_1,\vec n_2)=\sum_{M=m_1+m_2}C(jM;lm_1,lm_2)
Y_{lm_1}(\vec n_1)Y_{lm_2}(\vec n_2),
\ee
where $C$ are the Clebsch -- Gordon coefficients. This solution
describes the internal motion of the quantized string (glueball) with
spin $j$ and spin projection $M$ and pseudospins $l$, which take
values  $l=1,2,3,...$, in accord with stability of the classical
solution considered in the previous Section. The spin takes values
$j=0,1,2,...,2l$.

We see that the glueballs are space-parity even for all $j$ and $l$.
Since they are electrically neutral and charge-parity even, their
quantum numbers are
\be
I^Gj^{PC}=0^+j^{++}.
\ee

From Eqs. (46) and (47), the glueball mass depends on $l$ only.
Introducing  $k=0$ for the leading Regge trajectory, $k=1$ for the
first daughter trajectory, $k=2$ for the second one and so on, we can
write down for the spin
\be
j=2l-k.
\ee
Then from Eqs. (46), (47) and (49) the glueball Regge trajectories
are equal to
\be
\sqrt{(j+k)(j+k+2)}=a_0+\frac{1}{4\pi a}m^2.
\ee
Glueballs with even spins lie on the leading Regge trajectory, on the
second daughter trajectory and so on. Glueballs with odd spins lie on
the first, third daughter trajectories and so on. All these
trajectories $j(m^2,k)$ as functions of $m^2$ have at large $j$ the
slope
\be
j'(\infty,k)=\frac{1}{4\pi a}=0,452\pm0,005~Ē'^{-2},
\ee
which is twice as small as for quark-antiquark states and are
nonlinear at small $j$.

The glueball quantum numbers (50) and the mass degeneracy of the
states with different spins $j$ at the same pseudospin $l$ are
remarkable properties of glueballs in this model. The lightest states
with $l=1$ and quantum numbers $0^+0^{++}$, $0^+1^{++}$ and
$0^+2^{++}$ can be identified with the mesons
\begin{eqnarray}
f_0(1500),& 0^+0^{++},& m=1500\pm10 ~Β' \nonumber\\
f_1(1510),& 0^+1^{++},& m=1518\pm5  ~Β' \nonumber \\
f_2(1565),& 0^+2^{++},& m=1542\pm22 ~Β'. \nonumber\\
\end{eqnarray}
These mesons are not the quark-antiquark ones [1]. References to the 
original papers on these mesons and their discussions are given in
Ref. [5], where only $f_0(1500)$ is considered as firmly established. 
Let us conservatively estimate the lightest glueball mass with $l=1$
to be
\be
m_1=1500\pm20 ~Β'.
\ee
This alows one to obtain the constant $a_0$ from Eq. (52)
\be
a_0=1,81\pm0,04.
\ee
Now, when we know the model parameters we can fix the glueball Regge
trajectories (52) and to predict masses and quantum numbers of
heavier glueballs. For $l=2$ we have spin-parities and mass
\be
0^{++}, 1^{++}, 2^{++}, 3^{++}, 4^{++};  m_2=2610\pm20 ~Β'.
\ee
For the next glueballs the model predicts
\be
0^{++}, 1^{++},..., 5^{++}, 6^{++};  m_3=3360\pm25 ~Β'.
\ee

From Eq. (52) we get the glueball leading Regge trajectory --- the
Pomeron trajectory
\be
j(m^2, 0)\equiv j(m^2)=\sqrt{(a_0+\frac{1}{4\pi a}m^2)^2+1}-1,
\ee
with the intersept
\be
j(0)=1,07\pm0,03,
\ee
which corresponds to the high-energy data on hadron scattering [6]. 
Let us also note that
\be
j'(0)=0,395\pm0,005~ Ē'^{-2}, ~~~j''(0)\approx0.02 ~Ē'^{-4}.
\ee

Let us remark in conclusion that glueball decays can be considered
within this approach if interacting meson fields are introduced
instead of meson wave functions (second quantization), as it was done
in Ref. [7] for open strings. Thus, the string quark model provides
a single relativistic approach for description of quark-antiquark and
glueball meson states including prediction of the pomeron Regge
trajectory, with wide area of critical comparison with experiment.

From viewpoint of this model it is important to make more precise the
experimental status and properties of the  $f_1(1510)$ and
$f_2(1565)$-mesons and to obtain experimental information on possible
glueballs in the mass region of 2600 and 3300 $MeV$.

The author is grateful to V. A. Petrov for stimulating discussions.

\end{document}